\documentclass[preprint,aps,amssymb,amsmath,floats,showpacs,superscriptaddress]{revtex4}
\usepackage{graphicx}

\begin{document}
\title{The interplay of university and industry through the FP5 network}
\author{Juan A. Almendral}
\affiliation{Departamento de F\'{\i}sica, Universidad Rey Juan Carlos, Tulip\'an s/n, 28933 M\'ostoles, Madrid, Spain}
\author{J. G. Oliveira}
\affiliation{Departamento de F\'{\i}sica, Universidade de Aveiro, Campus Universit\'ario de Santiago, 3810-193 Aveiro,
Portugal}
\author{L. L\'{o}pez}
\affiliation{Departamento de Ingenier\'{\i}a Telem\'{a}tica y Tecnolog\'{\i}a Electr\'{o}nica, Universidad Rey Juan
Carlos, Tulip\'{a}n s/n, 28933 M\'{o}stoles, Madrid, Spain.}
\author{Miguel A. F. Sanju\'{a}n}
\affiliation{Departamento de F\'{\i}sica, Universidad Rey Juan Carlos, Tulip\'an s/n, 28933 M\'ostoles, Madrid, Spain}
\author{J. F. F. Mendes}
\affiliation{Departamento de F\'{\i}sica, Universidade de Aveiro, Campus Universit\'ario de Santiago, 3810-193 Aveiro,
Portugal}

\begin{abstract}

To improve the quality of life in a modern society it is essential to reduce the distance between basic research and
applications, whose crucial roles in shaping today's society prompt us to seek their understanding. Existing
studies on this subject, however, have neglected the network character of the interaction between university and industry. Here we use state-of-the-art
network theory methods to analyze this interplay in the so-called Framework Programme---an initiative which sets out
the priorities for the European Union's research and technological development. In particular we study in the 5th Framework Programme (FP5) the role played by companies
and scientific institutions and how they contribute to enhance the relationship between research and industry. Our
approach provides quantitative evidence that while firms are size hierarchically organized, universities and
research organizations keep the network from falling into pieces, paving the way for an effective knowledge transfer.

\end{abstract}
\date{\today}
\maketitle

{\bf We address, from a Complex Networks viewpoint, the relationship between companies and scientific institutions and
how they contribute to reduce the distance between research and applications. Since this approach requires information
about real systems, we focus on the Framework Programme (FP)---an initiative which sets out the priorities for the
European Union's research and technological development. Despite the presence of many and different participants, they
can be split in basically two groups: Companies and Universities. The first one is made of companies or other industry
related participants who expect that their investments in R+D+I are profitable. The second group can be thought as the
opposite: participants involved in some type of academic research for whom results do not necessarily return income.
We find that the transmission of information is more efficient between Universities than among Companies. Furthermore,
when Universities are excluded from the projects, Companies tend to form clusters, turning difficult (if not
impossible) the communication between them. Likewise, if we pay attention to the evolution of the FP, we see that,
although the creation of collaborations is encouraged, it is mainly between Universities and this is insufficient to
improve the relationship between research and industry. Finally, we find that Companies and Universities differ in the
way they establish collaborations. Large corporations are reluctant to choose as partners small companies, whereas size
is not important between Universities. But if we analyze how Universities and Companies cooperate, the result is that
large Universities prefer working with large Companies, while Companies select their collaborators between Universities
regardless of their sizes. These findings have potential implications for future programmes, as well as for new
policies and services aiming at research, development and innovation in general.}

\section{Introduction}

Understanding the relationship between research and industry is essential to improve the quality of life in any modern
society. Ranging from faster application of new discoveries to knowing whether or where investment should be employed,
this flow of knowledge between research and industry has long been of general interest. Yet, knowledge is a very
special resource whose study demands new techniques. The traditional approach to resources is based on the concept of
scarcity since they are usually finite. But knowledge cannot be seen this way because it grows, and the more it is used
the more it spreads~\cite{amidon}. In addition, existing studies on the research and industry
interplay~\cite{branscomb,caloghirou,krahmer} have neglected its network character. Our approach consists in analyzing
this issue from a complex network viewpoint~\cite{albert02}. Many other systems are better understood in this
manner~\cite{guimera05,jeong00,mendes03,newman04sci}. In this approach, the interaction between research and industry
is best described as a network whose vertices (or nodes) represent either companies or institutions devoted to research, and each
edge (or link) represents collaboration between any two of them. Hence, we can quantitatively study how research and industry
influence each other, if we have access to data describing a real system.

Here, we focus our attention in the so-called Framework Programme (FP), a mechanism aiming to improve the transference
of knowledge in the European Union (EU) by setting out its priorities for research and technological development. The
data to generate the corresponding FP network were gathered from the CORDIS website~\cite{cordis} by a robot. Since, currently, the
6th programme is under execution and the 7th is being planned, we focused our study in the 5th Framework
Programme (FP5)---covering the period from 1998 to 2002---in order to analyze a completely finished programme. Despite the presence of more than $25,\!000$
participants, they can be split in two major groups: Companies and Universities. The first is made of over $16,\!700$
companies and other industry related participants who expect their investments in R+D+I to be profitable. The second
group can be regarded as the opposite, more than $8,\!500$ participants involved in some type of research for whom
results do not necessarily return immediate income (see Appendix). Exploring the relationship between these two groups not only
provides a good example of the interplay between structure and information flow, but also offers a glimpse on how
research links with innovation and if the distance between basic research, applications and products
reduces~\cite{wigzell}.

It is worth remarking that we are mainly interested in the capacity of the FP5 to create and transfer
information and nothing can be said about this issue inside each node. Notice that some participants are large
institutions or companies with complex organization charts, which may have several projects whose coordination cannot
be guaranteed in general. However, our main concern is how to set the means to integrate research, development and
innovation efficiently, not if these means are successfully used.

\section{Analysis of the data}

To characterize the FP5, in this section we compute five important features in any network: degree distribution, shortest path
distribution, betweenness centrality, clustering coefficient and the degree--degree correlation. The detailed description of the
dataset can be found in the Appendix.

\subsection{Degree distribution}

\begin{figure}[htb]
\centering
\includegraphics[width=.45\textwidth]{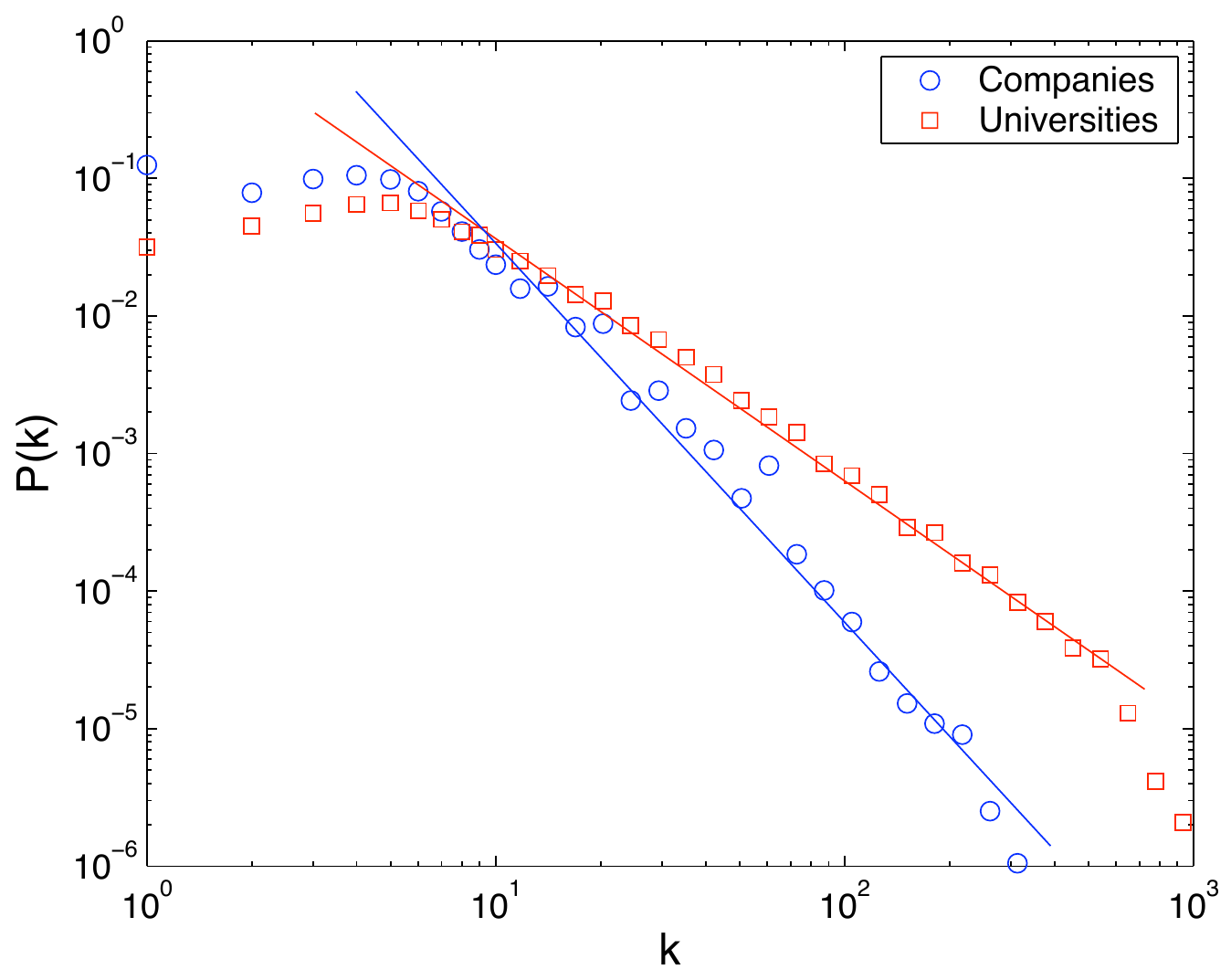}

\caption{This graph depicts with red squares the probability that a University collaborates with $k$
other Universities, that is, its degree distribution. The degree distribution of Companies is shown with blue circles. Data were log-binned.
We find that both distributions follow a power law tail, $P(k) \sim k^{-\gamma}$, thus having a scale--free topology, with vertices connecting each other in a heterogeneous manner: Most vertices have few connections, but some have a very large degree. The best fit for the straight region of the curves gives $\gamma_U=1.76 \pm
0.01$ with a correlation coefficient $R=0.998$ for Universities, and $\gamma_C=2.76 \pm 0.03$ with $R=0.991$ for
Companies. However, the fact that Universities show $\gamma_U <2$ whereas Companies have $\gamma_C >2$ implies that the
mean degree of Universities grows in time but not the mean degree of Companies. This result suggests that some
form of synergy encourages the creation of new collaborations mainly between Universities, while the network of Companies is less dynamic in this respect.}
\label{F:P(k)}
\end{figure}

The probability that a University collaborates with $k$ other Universities (i.e., the degree distribution of the
Universities) decays as a power law, $P(k) \sim k^{-\gamma_U}$ with $\gamma_U =1.76$. Similarly, Companies follow a
power law with $\gamma_C =2.76$. The two distributions can be seen in Fig.~\ref{F:P(k)}, where a log--log scale is
used in the plot, providing evidence for the scale--free topology~\cite{barabasi99} of both networks. The degree distribution of the whole FP5 network is also well approximated by a power law with exponent $\gamma$ close to $2.1$ \cite{almendralPhysA}.

Note that the degree distribution of Universities is described by a power law with $\gamma_U<2$, implying that their
mean degree grows in time. Indeed the first moment (i.e. mean degree in this case) of a distribution with a power--law tail diverges when its exponent is less than 2. This result suggests that Universities form an accelerated growing network~\cite{mendes02}, where the total number of edges grows faster than a linear function of the total number of vertices and,
consequently, it is verified that $1< \gamma <2$.

To elucidate this issue, we computed the average degree $\langle k \rangle$ during several years to check its tendency.
Though we only have the data corresponding to $4$ years (table~\ref{table}), they are enough to confirm the existence
of an accelerated growth since the average degree is not constant (46\% increase for the network of Universities in the four year period). But if the collaborations grow faster than
proportional to the number of participants, it is because they do not emerge by the mere increase of participants. Not
only new participants contribute to increase the number of collaborations, but also the old ones, meaning that some
form of synergy exists encouraging the creation of new collaborations between Universities.

On the other hand, the average degree of Companies also grows (though significantly slower)
during the four year span of the dataset
(table~\ref{table}). However, the fact that $\gamma_C>2$ suggests that this increase should be transient. Therefore, although the creation of collaborations is encouraged (since when the FP5 was finished
the mean number of collaborations had risen from $10$ to $26$ and some participants had surpassed $2,\!500$
collaborations) these results reveal that the synergy is more pronounced between Universities. In this sense, the FP5 is less effective in improving the network of Companies and Universities seem to take more advantage of this opportunity to create new collaborations.

Also noticeable in table~\ref{table} is the fact that the number of Companies increases faster than the number of Universities (72\% and 64\% increase respectively in the four year period), indicating another difference in the evolution of both networks.

\begin{table}[htb]
\centering
\begin{tabular}{|c|c|c|c|}
    \hline      & $N$             & $\langle k \rangle$   & $\langle C \rangle$   \\
           Year & Univ--Comp      & Univ--Comp              & Univ--Comp          \\
    \hline 1999 & 3075--4658    & 17.2--6.2             & 0.65--0.58            \\
    \hline 2000 & 5377--9359    & 21.9--6.8             & 0.66--0.53            \\
    \hline 2001 & 7355--13905   & 27.7--7.9             & 0.67--0.53            \\
    \hline 2002 & 8522--16765   & 31.9--8.2             & 0.68--0.59            \\
    \hline
\end{tabular}

\caption{Evolution of Universities and Companies during the FP5. Here we show the total number of vertices $N$, the
average degree $\langle k \rangle$ and the average clustering coefficient $\langle C \rangle$ during the four years
that the FP5 lasted.} \label{table}
\end{table}

\subsection{Shortest paths}

\begin{figure}[htb]
\centering
\includegraphics[width=.45\textwidth]{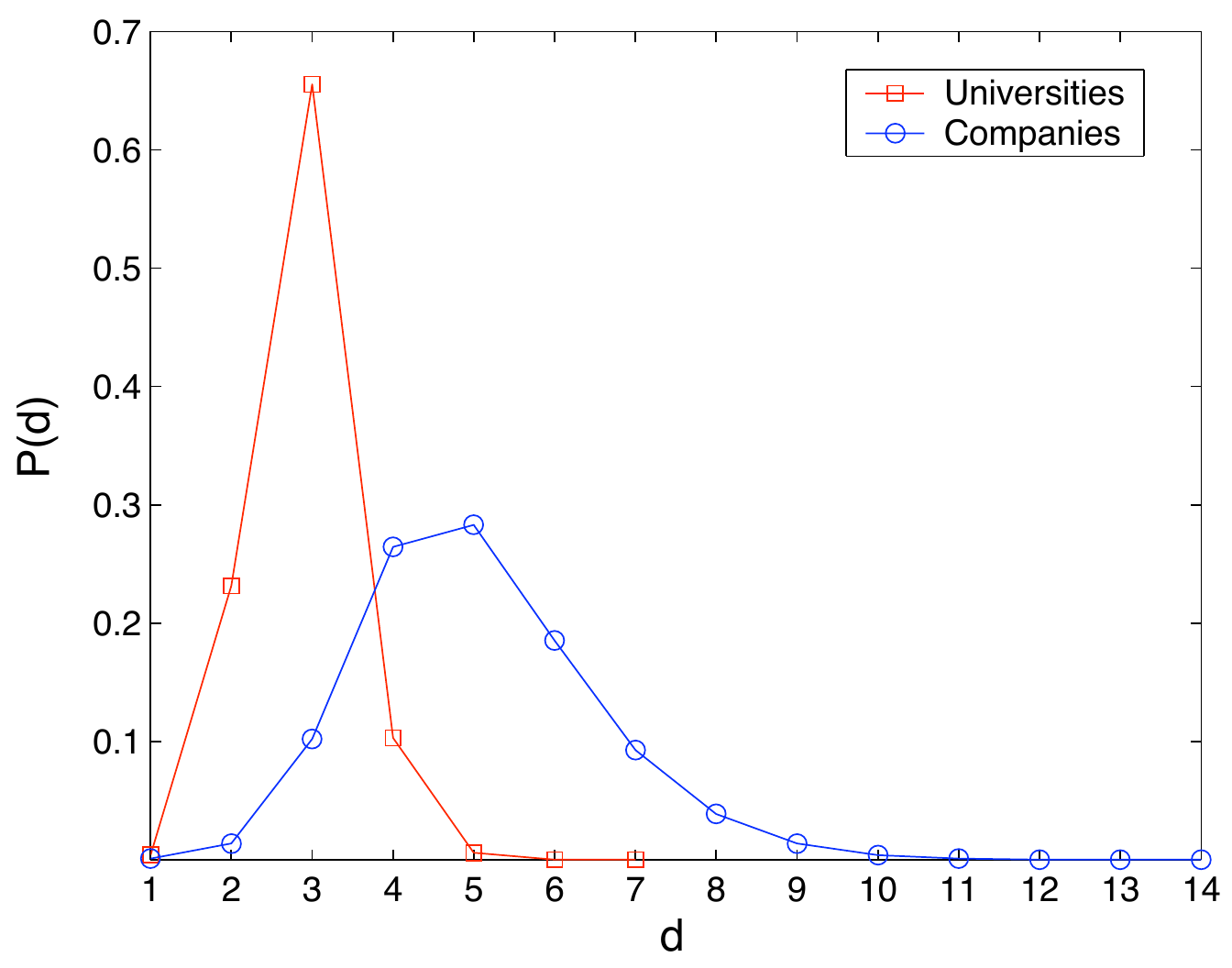}

\caption{The distribution of shortest paths in the largest connected component of Universities (red
squares) and Companies (blue circles) displays the presence of the small--world effect. The mean value is $\langle d \rangle =3.34$ for Universities and
$\langle d \rangle =5.67$ for Companies. Moreover, while the farthest pair of Companies has $13$ intermediaries, for
Universities the maximum separation is $7$ edges. Therefore, Universities are important for Companies since, when they
cooperate, in the whole FP5 network the largest distance reduces to $8$ and the average distance to $3.14$.}
\label{F:P(l)}
\end{figure}

A {\em path} between two participants is defined as a sequence of edges which links them, the distance between them being the
number of edges in the shortest path. Defining the set of participants which can be linked through a path as a
connected component, we find that the largest connected component of
Universities spans $93.7\%$ of the network ($7,\!987$ vertices) while for Companies it is made of $10,\!801$ nodes
($64.4\%$). Hence, while almost all Universities are linked in only one component, Companies are more fragmented and one third of
them fall in other smaller components (actually, the second biggest component contains only $48$ participants). This
result shows that Universities are important to compact the network since the largest connected component of the complete network (U+C)
comprises $88.7\%$ of the Companies and $96.0\%$ of the Universities (i.e. $23,\!055$ vertices in total). In addition, the largest
distance in the network of Universities is $7$ and the average distance is $\langle d \rangle =3.34$ whereas, in the
case of Companies, the farthest pair is separated by $14$ edges and the average distance is $\langle d \rangle
=5.67$~\cite{randgraph}. This can be seen in Fig.~\ref{F:P(l)} where we plot the distance distribution, $P(d)$ versus
$d$. Hence, also here Universities are essential for Companies since the largest distance in the entire network is only
$8$ and the average distance is $\langle d \rangle =3.14$, which implies that, on average, there are only two
intermediaries between two participants.

\begin{figure}[htb]
\centering
\includegraphics[width=.45\textwidth]{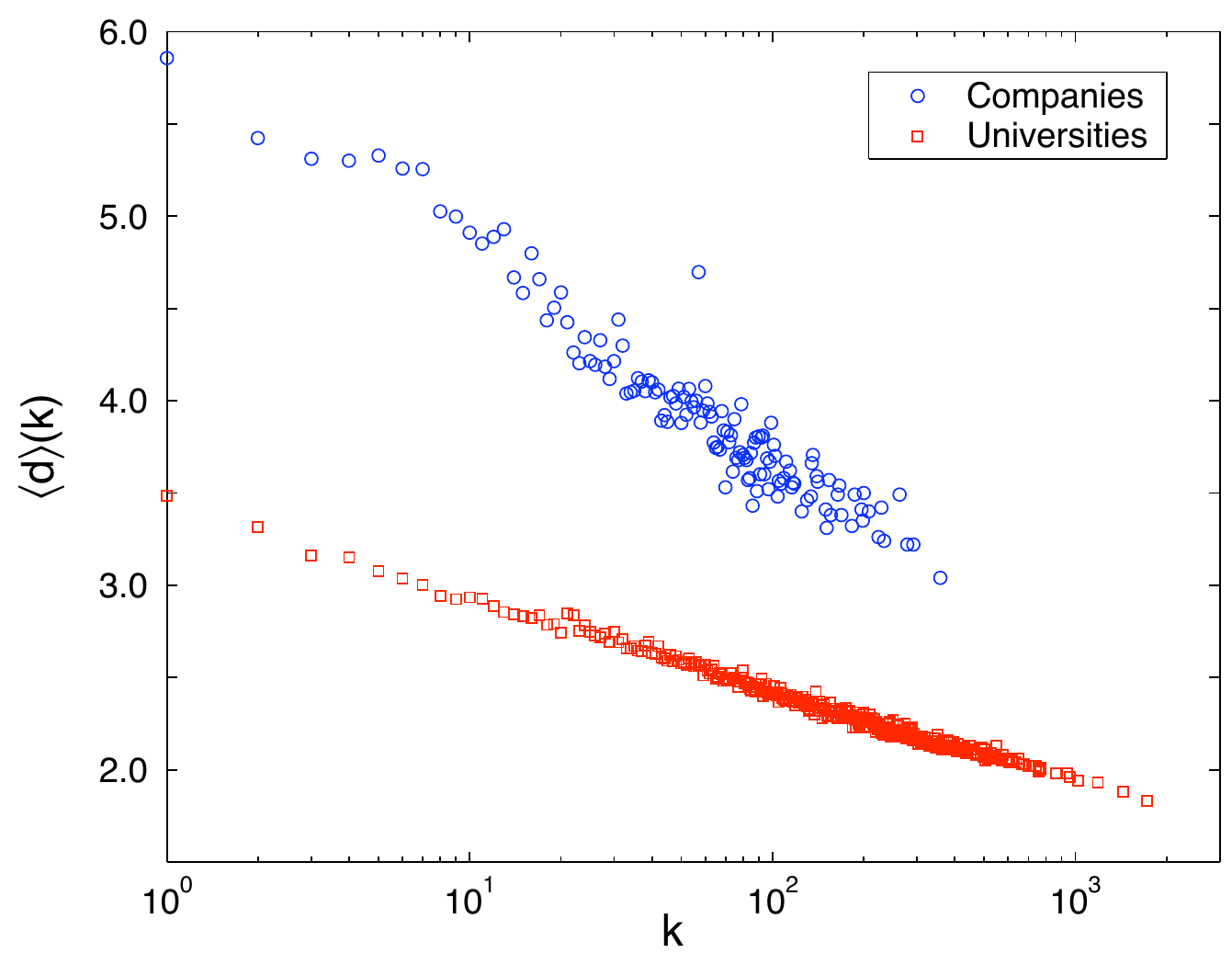}

\caption{The average distance of a participant with $k$ partners to all other participants in the largest
connected component is depicted. Universities are the red squares and Companies are the blue circles. It can be seen
the logarithmic dependence since it is verified that $\langle d \rangle \sim - \beta \log k$ where $\beta_U =0.503 \pm
0.003$ with $R=0.994$ for Universities and $\beta_C =1.13 \pm 0.03$ with $R=0.958$ for Companies. The decay is faster (i.e. $\beta_C > \beta_U$) in the net with the larger value of exponent $\gamma$ (see Fig.~\ref{F:P(k)}), providing empirical  evidence for the results of Ref.~\cite{oliveira04}. Note that the lowest
degree vertices in the network of Universities show a distance to other vertices comparable to the one of the highest
degree vertices in the network of Companies. Also note that in both networks $\max \d(k) \approx 2\min  \d(k)$ as had been previously observed in network models~\cite{oliveira04}.} \label{F:l(k)}
\end{figure}

The average distance is a coarse characteristic though. As a finer measure, it is possible to compute the average distance of a vertex of degree $k$ to all other vertices in the
largest component~\cite{oliveira04}. In Fig.~\ref{F:l(k)} we plot $\langle d \rangle (k)$ for both networks on a log--linear scale, where the Y axis means $\langle d \rangle (k)$ and the X axis is
$\log k$.

Therefore, albeit both networks display the so-called small--world effect~\cite{strogatz01}, there are important
differences. The presence of Universities eases the flow of information since they are much closer to each other than
Companies. This could be expected since the main purpose of a company is to satisfy its shareholders, which does not
include the spread of information from which competitors can take advantage. But, interestingly, the consequences of
this fact go beyond. When Universities are excluded from the projects, Companies become isolated despite Universities
are only one third of the participants. Companies tend to form clusters, turning difficult (if not impossible) the
communication between them and, consequently, little can be developed or innovated since other results are not
available to work with. Thus the natural tendency of Companies to protect their findings would finish killing R+D+I.
The presence of Universities contributes to moderate this.

\subsection{Betweenness centrality}

To further investigate the interplay between the two kinds of participants, we can also measure the
{\em betweenness centrality}~\cite{freeman77} in the FP5. The betweenness $\sigma_m$ of vertex $m$ measures the extent to which $m$ lies
on the paths between other participants. It therefore accounts for the influence of a participant between other two
distant participants, relating the local structure and the global topology of the network. It is
defined as
\[ \sigma_m = \frac{1}{(N-1)(N-2)} \sum_{i,j:i \ne j \ne m} \frac{B(i,m,j)}{B(i,j)}, \]
where $B(i,j)$ is the number of shortest paths between nodes $i$ and $j$, $B(i,m,j)$ is the number of such shortest
paths passing through vertex $m$, and the sum is taken over all pairs of vertices $i$ and $j$ which do not include $m$.
The pre-factor, where $N$ is the total number of nodes, accounts for normalization, so that $0 \leq \sigma_m \leq 1$.

Since the computation of the betweenness for the whole FP5 is an extremely time-consuming task, we focus our study on
one of its subprograms: `Promotion of innovation and encouragement of small and medium sized enterprises participation' (SME), which is formed by $195$ research institutions and
$212$ Companies (see Appendix). Given our ability to split the SME into Universities and Companies, several different
situations are considered. The average betweenness of the SME, taken over all its vertices, turns out to be $\langle
\sigma \rangle = 5.19 \cdot 10^{-3}$. Considering only those vertices $m$ which are Universities, we find that their
average betweenness among all other vertices in the SME is $\langle \sigma_U \rangle = 6.76 \cdot 10^{-3}$. Likewise,
we obtain $\langle \sigma_C \rangle = 3.74 \cdot 10^{-3}$ for Companies.

Now, if we only take into account those shortest paths whose endpoints are Companies, the betweenness measures the role
Universities play in linking Companies: $\langle \sigma_{CUC} \rangle = 5.44 \cdot 10^{-3}$; on the other hand, when
the endpoints are Universities, the average betweenness of Companies is $\langle \sigma_{UCU} \rangle = 2.34 \cdot
10^{-3}$. Thus, we see that the role Universities play between Companies is more than twice the one played by
Companies between Universities. Moreover, given that $\langle \sigma_U \rangle > \langle \sigma \rangle > \langle
\sigma_C \rangle$, we observe again the central function played by research institutions in the FP5 network.

\subsection{Clustering coefficient}

The {\em clustering coefficient} of a vertex $i$ is defined as $C_i = 2n_i/[k_i(k_i-1)]$, where $n_i$ is the number of edges
connecting its $k_i$ nearest neighbors. It equals $1$ for a participant at the center of a completely connected
cluster, and $0$ for a node whose neighbors are not linked at all. Taking the average of the clustering coefficient, we
obtain $\langle C \rangle = 0.68$ for Universities and $\langle C \rangle = 0.59$ for Companies, which are much higher
than the average clustering coefficient of a random graph~\cite{bollobas85} with the same number of nodes and average
degree (namely, $\langle C \rangle = \langle k \rangle/N$). Moreover, $\langle C \rangle$ is independent of the number
$N$ of participants in both cases (see table~\ref{table}), in contrast with the prediction of a scale--free
model~\cite{barabasi99} where $\langle C \rangle \sim N^{-0.75}$~\cite{albert02,bollobas03}. This high and size--independent
average clustering coefficient evidences the organization of Universities and Companies in modules.

\begin{figure}[htb]
\centering
\includegraphics[width=.45\textwidth]{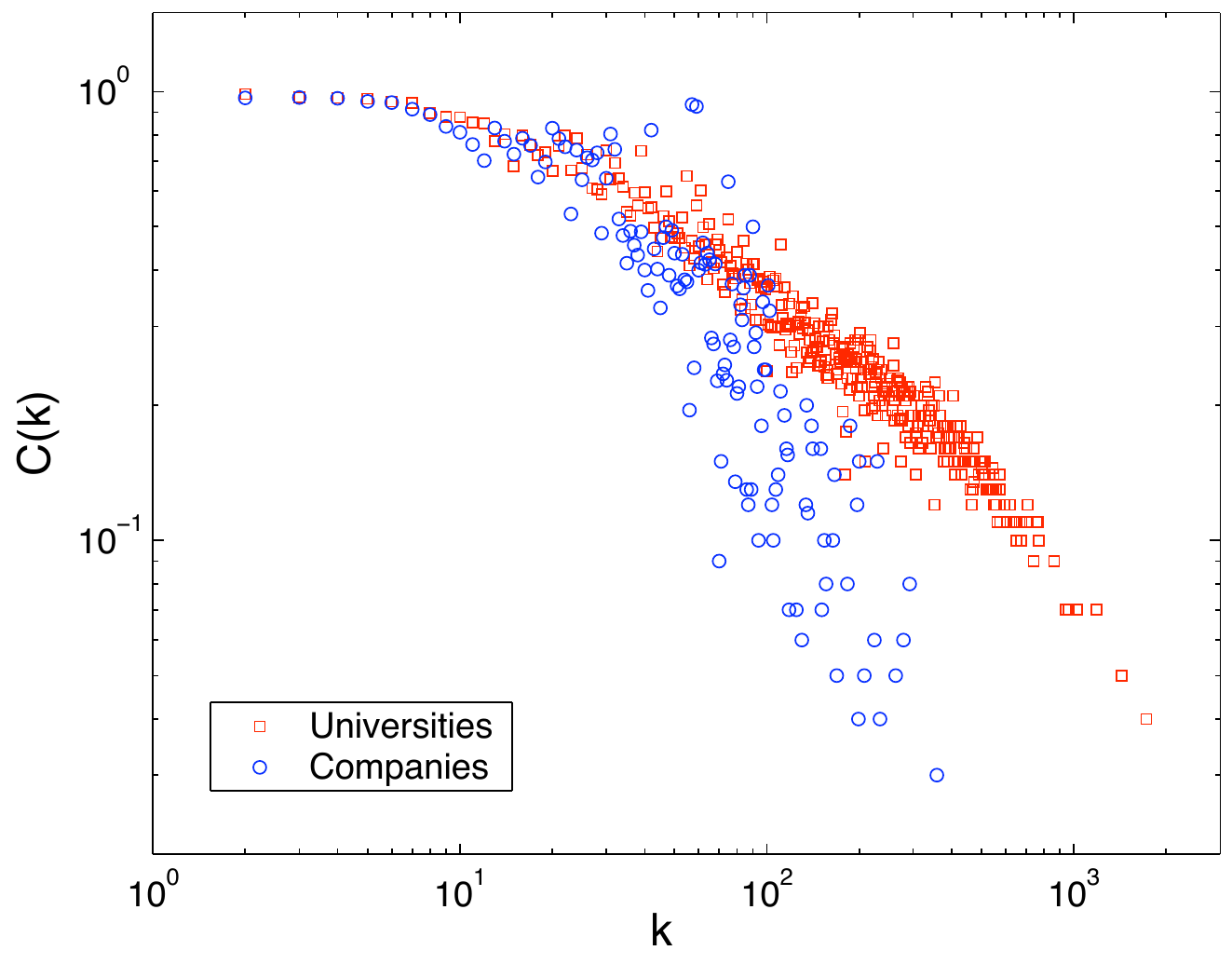}

\caption{In this graph the clustering coefficient as a function of $k$ is shown. After the initial
plateau, where $C(k)$ is approximately constant, it decays as a power law, $C(k) \sim k^{-\alpha}$, where $\alpha_U =
0.54 \pm 0.01$ with $R=0.97$ for Universities (red squares) and $\alpha_C = 1.05 \pm 0.06$ with $R=0.86$ for Companies
(blue circles). We therefore conclude that both networks have hierarchical modularity since scale--free and modular
networks are degree--independent, whereas hierarchical modularity is characterized by the power-law decay $C(k) \sim
k^{-\alpha}$.} \label{F:C(k)}
\end{figure}

However, when we measure the clustering coefficient of a node with $k$ links, $C(k)$, for both networks
(Fig.~\ref{F:C(k)}), we find that it decays as a power law for large $k$. We therefore infer that the two nets have
hierarchical modularity, which is characterized by the scaling law $C(k) \sim k^{-\alpha}$, in contrast to some scale--free or modular networks where the clustering coefficient is degree--independent~\cite{barabasi02}.

\begin{figure}[htb]
\centering
\includegraphics[width=.45\textwidth]{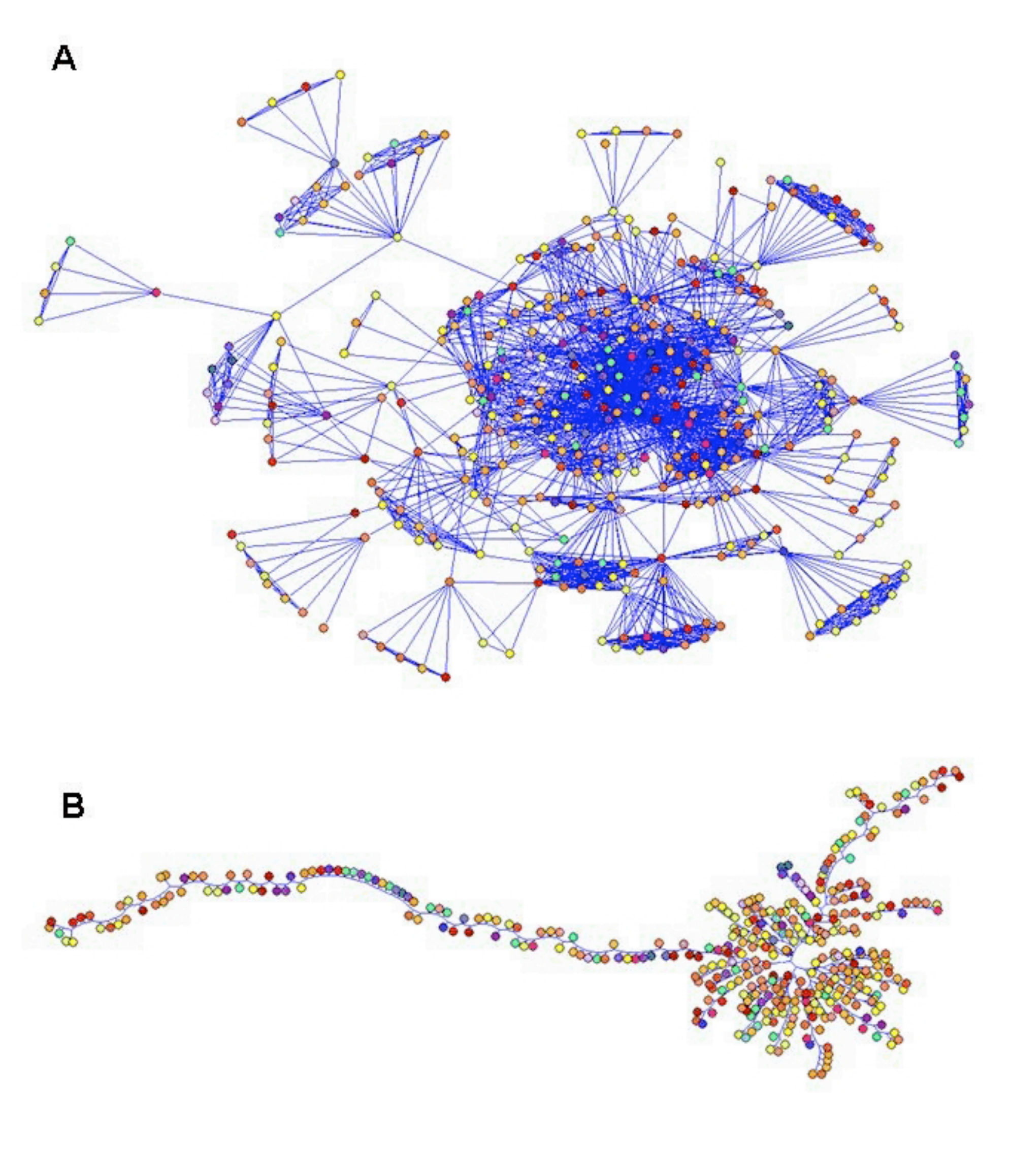}

\caption{The existence of hierarchical modularity in the networks of Universities and Companies suggests
that they have a self--similar structure. Since projects in the FP are classified in $8$ subprograms depending on their
objectives, we choose, for clarity, to illustrate in {\bf A} this self--similar structure with the smallest one: `Promotion of innovation 
and encouragement of small and medium sized enterprises participation' (SME). Also, to verify if there is a bias by nationality in the collaborations, we searched
for communities reflecting groups of participants collaborating strongly among them. In the networks of Universities,
Companies and both together (even when they are analyzed by subprogram) the result was similar to {\bf B,} corresponding to
the SME subprogram. If we color the nodes according to their nationalities and arrange them in space with a standard
algorithm~\cite{pajek}, we find that they are all mixed.} \label{F:pajek}
\end{figure}

This result suggests that Universities and Companies have an inherent self--similar structure~\cite{makse05}, being
made of many highly connected small modules, which integrate into larger modules, which in turn group into even larger
modules (Fig.~\ref{F:pajek}A). Actually, we observe that $4,\!333$ Universities ($50.8\%$) and $10,\!564$ Companies
($63.5\%$) have $C_i =1$, indicating the presence of many totally connected groups. This is due to the fact that most
of these entities participate in only one project, having as neighbors other vertices, which in turn are all connected
between them by virtue of the participation in the project. Furthermore, given that this result suggests
weak geographical constraints~\cite{barabasi03}, we searched for communities in them~\cite{newman04pre} and found
precisely that they were not based on nationality (Fig.~\ref{F:pajek}B), whence, the FP is successfully applying a
policy which avoids its segregation by nationality.

\subsection{Degree--degree correlation}

An interesting question is which vertices pair up with which others. It may happen that vertices connect randomly, no
matter how different they are. Usually, however, there is a selective linking, i.e. there is some feature which
makes more (or less) likely the connection~\cite{newman03rev}. There is {\em assortative mixing} when 
vertices of similar degree tend to be connected, and {\em disassortative mixing} in the opposite case (i.e. when vertices of high degree tend to connect to vertices of low degree)~\cite{newman02,newman03pre}.

A first approach to elucidate this issue is by means of the {\em joint degree--degree distribution} $P(k,k')$, which gives us
the probability of finding an edge connecting vertices of degree $k$ and $k'$. We see that for Companies the distribution has sharp peaks
for $k=k'$ (Fig.~\ref{F:P(k,k')}A). This network thus seems to display assortative mixing, i.e. if one chooses at random a vertex of degree $k$ then, with
considerable probability, it will be connected to vertices of degree $k$. In other words, Companies with similar degree tend to collaborate more frequently than Companies with different
degrees.

Notice that the fact (mentioned in the previous section) that many entities participate in only one project may, by itself, explain these peaks: If the $X$ participants of a certain project have no other
projects each of them has degree $X-1$ and each of their neighbors has degree
$X-1$, giving rise to an assortative trend. On the other hand one can also argue that, when a Company has high degree it is due to being involved in many projects. It is then
reasonable to assume that nodes with high degree represent large institutions, given that only these can deal with many
projects at the same time. That being the case, the observed assortativity means that the spread of information between Companies
depends on the institution's size. On the contrary, for Universities $P(k,k')$ is scattered throughout the plane $k-k'$
(Fig.~\ref{F:P(k,k')}B). While there are still peaks along the line $k=k'$, the presence of many others for $k \ne k'$
is clear, suggesting that Universities are less selective in what regards the size of their partners.

\begin{figure}[htb]
\centering
\includegraphics[width=.45\textwidth]{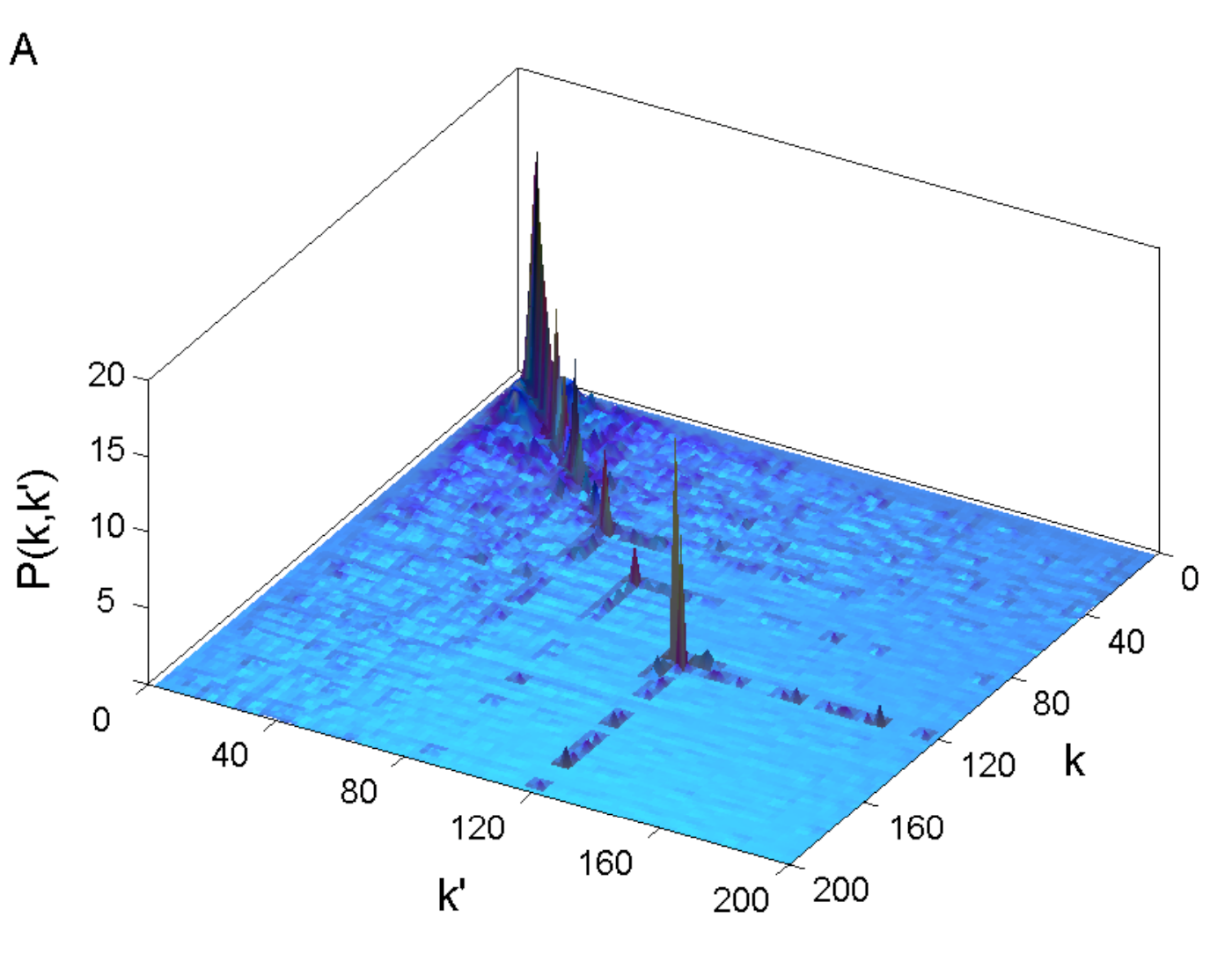}
\includegraphics[width=.45\textwidth]{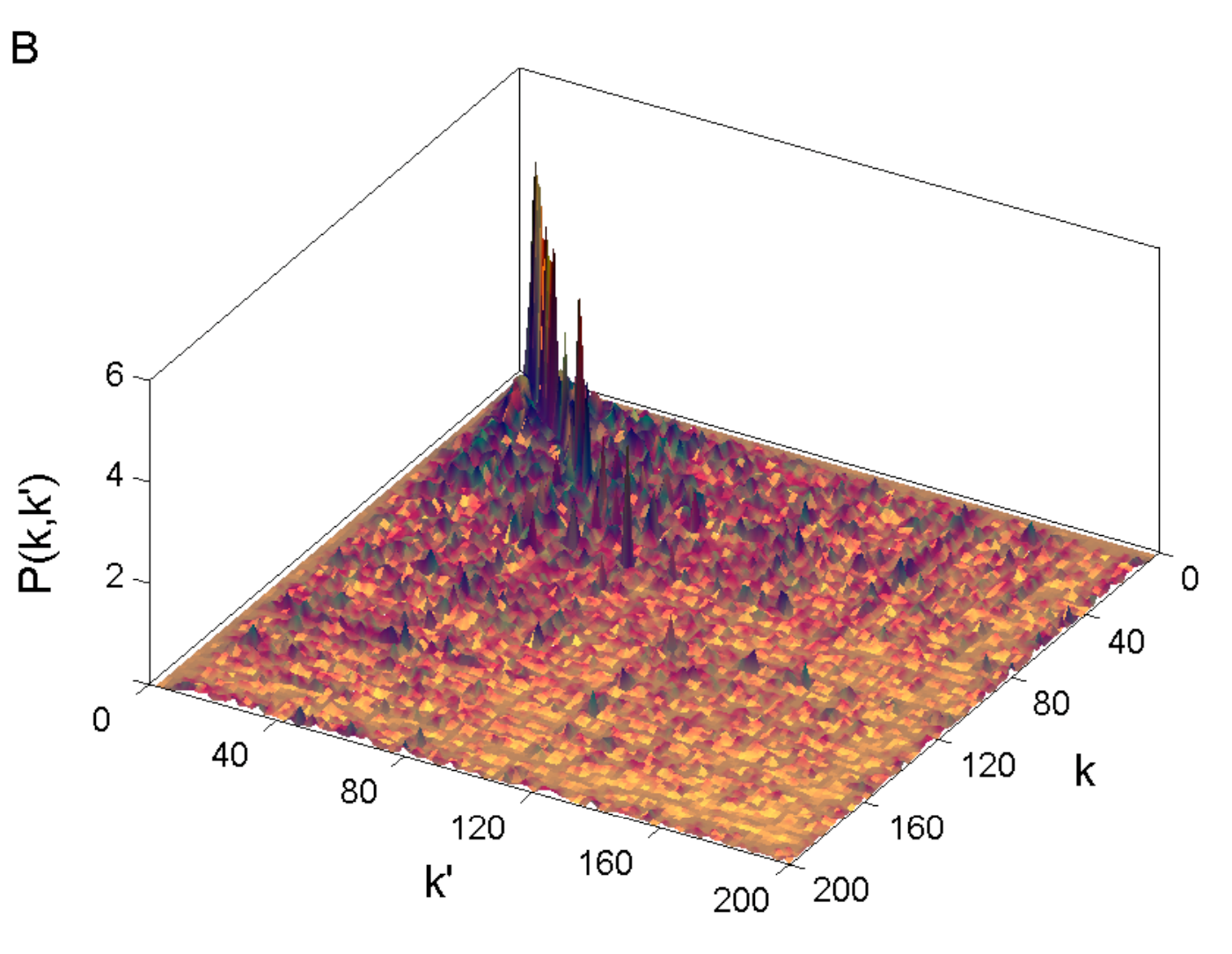}

\caption{Determination of the mixing through the joint degree--degree distribution. The X and Y axes
represent the degrees $k$ and $k'$ and the Z axis gives the corresponding joint degree--degree probability in per mill.
The range is limited from $0$ to $200$ to illustrate a clearer picture. The joint degree-degree distribution of
Companies ({\bf A}) peaks on the line $k=k'$ which implies that the mixing is assortative. Since the number of links held by
a participant is related to its size, we infer that Companies with similar sizes tend to collaborate more frequently
than Companies with different sizes. The joint degree--degree distribution of Universities ({\bf B}) is distributed
throughout the X--Y plane which suggests that Universities do not have assortative mixing and thus choose their collaborators in a less selective manner.} \label{F:P(k,k')}
\end{figure}

\begin{figure}[htb]
\centering
\includegraphics[width=.45\textwidth]{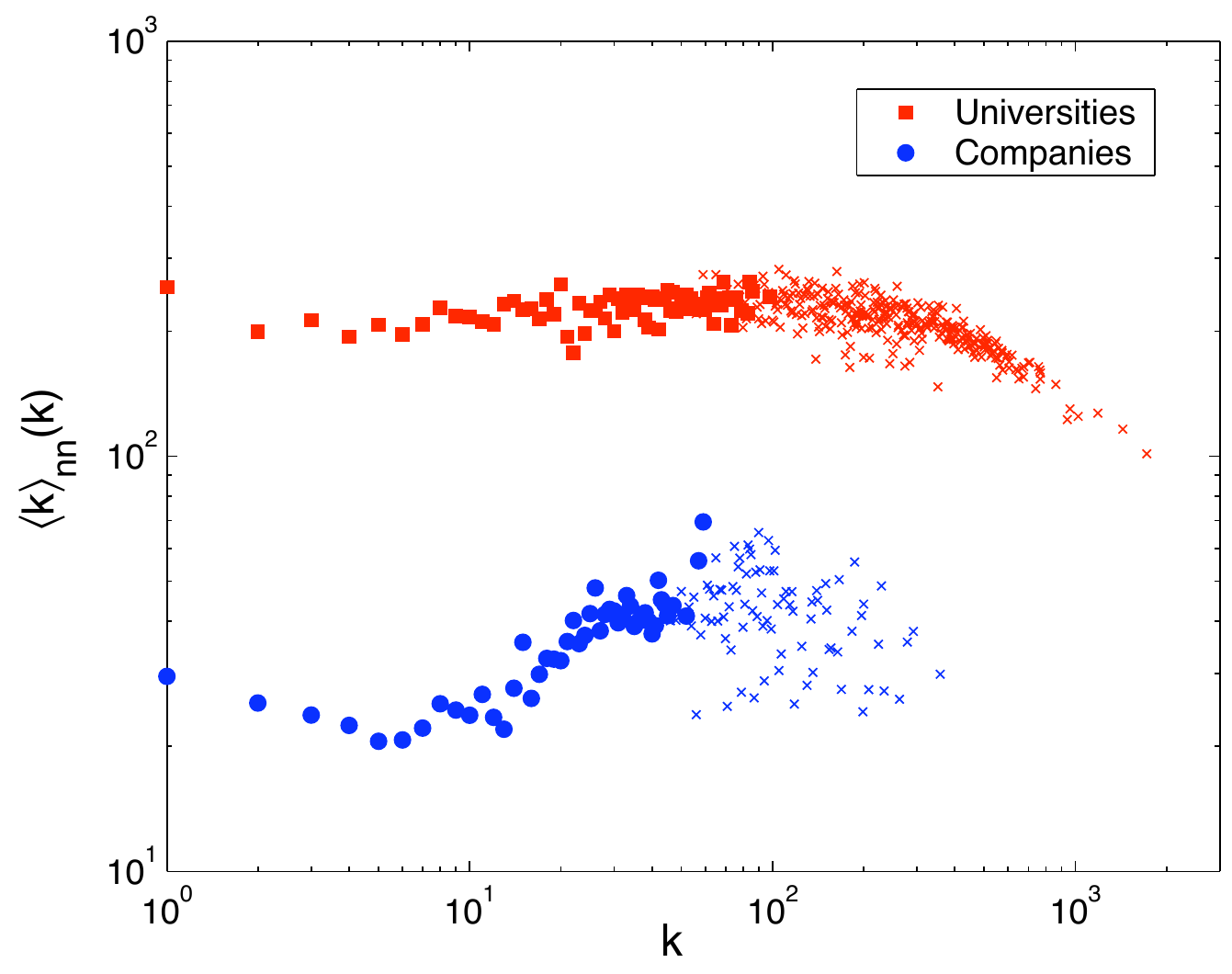}

\caption{In this plot the average degree of the nearest neighbors of a vertex with $k$ links, $\langle k
\rangle_{nn}(k)$, is shown. To mark the proximity to the cut--off, the points obtained from less than $10$ observations
are plotted as crosses (Universities in red and Companies in blue) and the remaining points as squares
(Universities) or circles (Companies). In this manner, it can be seen that these points are biased downwards due to
the finite size of the network. Then, once focusing our attention on the circles and the squares, we find that
Companies have assortative mixing, while Universities link between them regardless their degrees.} \label{F:knn}
\end{figure}

It is important to remark, however, that the joint degree--degree distribution requires many observations in order to obtain
good statistics. For example, if we focus our analysis in the range $[0,200]$, we need about $200 \times 200$ points,
otherwise fluctuations are important and the plot is far from smooth~\cite{boguna04}. To avoid this problem, one uses
the average degree of the nearest neighbors of a vertex of degree $k$, $\langle k \rangle_{nn}(k)$, which is a coarser
but less fluctuating measure. To compute it, we find all participants with $k$ links and take the average degree of all
their neighbors. The results are shown in Fig.~\ref{F:knn}, and confirm those obtained through the joint degree--degree
distributions. To emphasize the presence of the cut--off due to the finite size of the network, the points obtained
from less than $10$ observations are plotted as crosses (Universities in red and Companies in blue) and the rest of the
points as squares (Universities) or circles (Companies). Considering then only the circles and the squares, we
confirm that collaborations between Companies are size--dependent (positive slope) whereas those between Universities
are not (no slope).

\begin{figure}[htb]
\centering
\includegraphics[width=.45\textwidth]{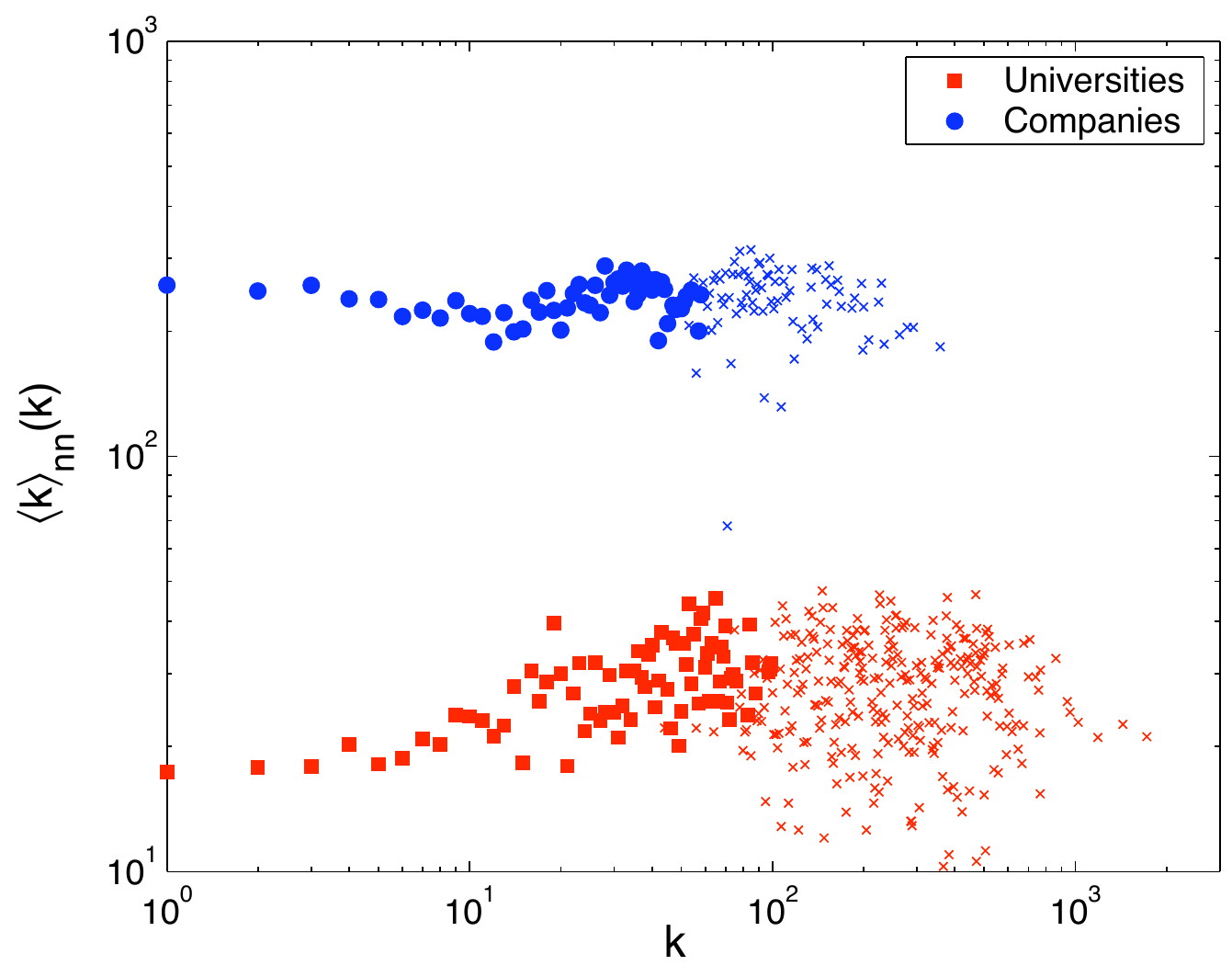}

\caption{Here we plot the average degree of the nearest Companies of a University with $k$ links to
other Universities (red squares) and the average degree of the nearest Universities of a Company with $k$ links to
other Companies (blue circles). As before, if we only consider the circles and the squares, we find that Companies link
to Universities independently of their degrees while Universities with high degree collaborate mainly with Companies
which have also high degree.} \label{F:knn_cross}
\end{figure}

It is also interesting to analyze how Universities and Companies link each other, which can be done as follows. We
search for all Companies with $k$ links and then compute the average degree of all their neighboring Universities.
Note that the former degrees are always calculated in the corresponding network, thus a Company with degree $k$ has
$k$ neighbor Companies, though it may have more links (to Universities) in the complete FP5 network. Analogously, we can find all Universities
with $k$ links to average the degrees of all neighbor Companies. The results are depicted in Fig.~\ref{F:knn_cross}
where, as before, it is used a $\log-\log$ scale. Again, we plot as squares (Universities) or circles (Companies) the
points obtained from more than $10$ observations to identify the region where the tendency is well defined. We find
that, while Companies link to Universities independently of their sizes, Universities with high degree tend to
collaborate with large Companies.

Finally, another way to quantify the mixing in the FP5 is by means of the {\em assortativity
coefficient}~\cite{newman02}, which is just the Pearson correlation coefficient of the degrees of connected vertices. In this case, we obtain what type of mixing takes place in the network by means of a
single number instead of a distribution. If $e_{jk}$ is the probability that a randomly chosen edge has vertices with
degree $j$ and $k$ at either end, the assortativity coefficient takes the following form:
\begin{equation*}
r = \frac{\sum_{jk} jk(e_{jk} -q_j q_k)}{\sum_k k^2 q_k - \left( \sum_k k q_k \right)^2}
\end{equation*}
where $q_k = \sum_j e_{jk}$ and $q_j = \sum_k e_{jk}$. This coefficient verifies that $-1 \le r \le 1$, being positive
when the network is assortative and negative when it is disassortative. We find $r_C=0.13$ for the network of Companies and $r_U=0.06$ for Universities, corroborating an assortative trend usual in social networks~\cite{newman03pre}.

Therefore, Companies and Universities differ in the way they establish collaborations. Companies are organized
hierarchically, where positions in that hierarchy are related to the size: The assortative trend in the network of Companies suggests that large corporations are reluctant to choose
as partners small companies. Between Universities, however, size is not important and it is common to find a large
institution collaborating with a small one. But if we analyze which partners Universities choose among Companies, we
check that large institutions in Universities prefer working with large Companies. On the contrary, Companies select
their collaborators between Universities regardless of their sizes. We can then conclude that large Companies play
indeed a leading role in the FP5 while Universities play the role of bridges between participants which are separated in
the hierarchical structure of Companies.

\section{Conclusion}

We have presented here a study of the interplay between research and industry in the scope of the Fifth Framework
Programme. Using network theory methods, we perform several measures that allow
us to quantify the features of this relationship and assess their potential improvements.
Naturally, the FP5 network does not include all interactions between university and industry (such as the recruitment of graduates by companies, the transfer of knowledge through scientific and technical literature or industry conferences). Furthermore, as already mentioned in the introduction, it also neglects the fact that internal connections in an institution (e.g. between different departments) may be absent, which would mean that a node in the studied network would split into disconnected nodes.
While these issues may significantly influence the flow of information in the network, addressing all of them requires information that is beyond reach for most researchers at this point.
The presented analysis thus represents a starting point for a quantitative understanding of the university--industry interplay network. It is possible, however, to foresee advances in these directions, given the increasing availability of information on how institutions self-organize.

The results point to the central function played by Universities in the FP5 network in reducing the distance between
research and applications. Indeed, we show that Universities play a crucial role in connecting the network of Companies, which would otherwise be separated in many small clusters.
While the network of Universities
is well integrated and established in accordance to what is observed for other social networks, the same doesn't seem
to apply for the Companies network, mainly due to its relatively small largest connected component. Competition is
probably the origin of this effect, which is moderated by the presence of Universities. It seems reasonable, then, to conclude that special attention should be devoted to company--company collaborations. Supporting this, is also the fact that new collaborations arise at a higher rate between Universities.

Our observations suggest in addition that Companies and Universities establish collaborations differently: While Companies seem to exhibit a hierarchical structure in terms of their size, Universities are less selective in their collaborations. We also observed that both networks display hierarchical modularity and that communities in the FP5 network are not nation-based. The FP appears then to mix all nationalities of the European Union, thus reaching one of its main goals: Promote the transfer of knowledge throughout Europe.

\section*{ACKNOWLEDGMENTS}
J.A.A., L.L., and M.A.F.S. acknowledge financial support from the Spanish Ministry of Science and Technology under project number BFM2003-03081 and project FIS2006-08525. J.G.O. acknowledges financial support from FCT Portugal (SFRH/BD/14168/2003). J.F.F.M. was
partially supported by projects POCTI (FAT/46241/2002 and MAT/46176/2002) and project DYSONET.

\section*{APPENDIX: CLASSIFICATION OF PARTICIPANTS INTO COMPANIES AND UNIVERSITIES}

The Framework Programme (FP) sets out the priorities for the European Union's research and technological development.
These priorities are defined following a set of criteria which pursue an increase of the industrial competitiveness and
the quality of life for European citizens. A fact which shows the effort made by the European Union to promote this
global policy for knowledge is the budget devoted to these programmes. For example, the FP5 (1998-2002) was implemented
by means of 13,700 million euros and the FP6 (2002-2006) has assigned a budget of 17,883 million euros.

All projects in the FP5 are organized in eight specific programmes which can be classified as follows. There are five
focused Thematic Programmes implementing research, technological development and demonstration activities:

\begin{itemize}
\item QOL: Quality of life and management of living resources (2,524 projects).

\item IST: User--friendly information society (2,382 projects).

\item GROWTH: Competitive and sustainable growth (2,014 projects).

\item EESD: Energy, environment and sustainable development (1,772 projects).

\item NUKE: Research and Training in the field of Nuclear Energy (1,032 projects).
\end{itemize}

And there are three Horizontal Programmes to cover the common needs across all research areas:

\begin{itemize}
\item INCO: Confirming the international role of Community research (1,034 projects).

\item SME: Promotion of Innovation and encouragement of small and medium enterprises participation (142 projects).

\item HPOT: Improving human research potential and the socio--economic knowledge base (4,876 projects).
\end{itemize}

The data to analyze the FP5 as a complex network were obtained from the web pages of CORDIS \cite{cordis} with a robot
implemented in Perl. The result was a database with 15,776 records as follows:

\vspace{10pt} \noindent Programme $|$ Year $|$ Participant1 - Nation - Dedication $|$ Participant2
- Nation - Dedication $|$ \dots…
\vspace{10pt}

The first field refers to the specific programme to which the project belongs and the second field informs us about the
year in which it started. The following fields are the participants in the project with their corresponding nationality
and dedication (`research', `education', `industry'...). We then have a bipartite graph \cite{albert02,mendes02} since
there are two kinds of vertices (participants and projects) and each edge links a participant with a project. To obtain
the graph with 25,287 participants (nodes) and 329,636 collaborations (edges) used throughout the text, we have only to
project it onto the participants.

The names of the participants were not free of typos since we collected them as they were in the web. The consequence
of this fact was that sometimes the same participant appeared in two projects with different names and, consequently,
it was recorded twice in the data. For instance, `Fran\c{c}ois Company of Something, Ltd.' and `Francois Company of
SOMETHING LTD' would be recorded as different. To avoid these duplications, we used a parser covering many
possibilities which could lead to false entries. Nevertheless, despite our efforts, not all duplications have been
eliminated. However, after a visual inspection of the data, we estimate that the error is below 10\%.

To split the participants in Universities and Companies, we considered the organization type reported in the project.
This information is encoded in the field `Dedication', where we found 11 levels: `Commission External Service',
`Commission Service', `Consultancy', `Education', `Industry', `Non Commercial', `Not Available', `Other', `Research',
`Technology Transfer' and $\langle\text{Void}\rangle$.

The level `Not available' means that the FP itself was not able to obtain the information and this absence is shown in
this manner. In addition, the level $\langle\text{Void}\rangle$ means that no information at all is given, i.e. our
robot found nothing (not even `Not Available').

The first step to define only two groups was to reduce the number of levels in `Dedication'. We found that eight levels
could be merged to define a new one, called `Non Companies'. It was not homogeneous since we found consultancies,
universities, hospitals, institutes, laboratories, observatories, museums, technological parks… even cities. However,
they all were participants involved in some type of research for whom results do not necessarily return income. This
new level was, basically, the union of `Research' and `Education' since the other six levels appeared few times in the
data: `Commission External Service' (4 records), `Commission Service' (8 records), `Consultancy' (49 records), `Non
Commercial' (389 records), `Technology Transfer' (1 record) and $\langle\text{Void}\rangle$ (1 record). The record with
$\langle\text{Void}\rangle$ was identified as `Non Company' by direct inspection.

Therefore, all records could be classified in one of the following levels: `Non Companies' (41,317), `Industry'
(6,447), `Other' (17,588) and `Not Available' (12,346). The total number of records (77,698) is larger than the number
of participants (25,287) since many of them collaborate in several projects. Then, it was necessary to verify if
repeated records were always classified in the same level of `Dedication'.

We found that many participants were classified in different levels, thus we had to define a set of rules which
eliminated this ambiguity. Hence, the following step was to study each level to understand their composition. For every
level, we chose 100 records randomly to check by direct inspection their dedication. The result was that all selected
records in `Industry' were companies, any in `Non Companies', 95 in `Other' and 55 in `Not Available'.

With the former information, we proceeded as follows. We first defined for each participant a vector D=\{`Non
Companies', `Industry', `Other', `Not Available'\}, where the components are the number of times that it is classified
in that level. For instance, D=\{17, 0, 8, 4\} means that the participant appears 17 times as `Non Company', 8 as
`Other' and 4 as `Not Available'. Then, we decided that vectors in the form \{a, 0, 0, 0\} or \{a, 0, 0, d\} were
Universities and vectors in the form \{0, b, c, d\}, \{0, b, c, 0\}, \{0, b, 0, d\} and \{0, b, 0, 0\} were Companies.
With only these sensible rules, we managed to classify 22,001 participants (87\%).

In order to confirm this result and to classify the remaining 3,286 entities, we defined a filter based in keywords
relative to the Universities group, such as `univer', `schule', `laborato'... When we focused our attention in the
group of 22,001 participants classified using `Dedication', we found that those classified as Universities according to
the filter were also Universities according to `Dedication'. Since the filter was a completely different manner of
splitting the dataset, we could use it for the rest of the entries. Note that we only believed the result of the filter
if it was University, not if the result was Company. This is reasonable since the filter was designed to identify terms
related to Universities, not to Companies.

By means of the filter we classified all participants but 309. To place these entities, we paid attention to which
value was higher: `Non Companies' or `Industry', independently of the other two values. If the value `Non Companies'
was higher, it was a University, otherwise it was a Company.


\begin{thebibliography}{9}
\bibitem{randgraph} Both average distances are approximately the value obtained for a random graph~\cite{bollobas85}
with the same number of nodes and average degree. For Universities is $\langle d \rangle \approx \log N/\log \langle k
\rangle =2.61$ and for Companies is $\langle d \rangle =4.62$.
\end{thebibliography}
\end{document}